# Tunable High-Quality Fano Resonance in Coupled Terahertz Whispering-Gallery-Mode Resonators

Shixing Yuan,[1] Liao Chen,[1] Ziwei Wang,[1] Ruolan Wang,[1] Xiaojun Wu,[2,a)] and Xinliang Zhang[1,b)]

[1]Wuhan National Laboratory for Optoelectronics and School of Optical and Electronic Information, Huazhong University of Science and Technology, Wuhan, 430074, China

[2]School of Electronic and Information Engineering, Beihang University, Beijing, 100191, China

Fano resonance is widely discussed in designing novel terahertz components, such as sensors, filters, modulators, and group delay modules. Usually, high quality ($Q$) factor and flexible tunability of Fano resonance are key requirements for these applications. Here, we present tunable terahertz Fano resonance with a $Q$ factor of 2095 at 0.439 THz in coupled terahertz whispering-gallery-mode resonators (WGMRs). Coupling between a relatively low $Q$ (578) quartz ring and a high $Q$ (2095) silicon ring is employed to generate the Fano resonance. The resonant frequency of the Fano resonance can be actively manipulated by tuning the resonant frequency of the high Q WGMR, which is achieved through utilizing an electrical thermo-optic tuning method, meanwhile, the resonance intensity of the Fano resonance can be engineered by adjusting the coupling strength between two WGMRs. This coupled-WGMR scheme delivers high Q tunable Fano resonance and may contribute to the design of high-performance configurable terahertz devices.

Electronic mail: a) xiaojunwu@buaa.edu.cn; and b) xlzhang@mail.hust.edu.cn.

Fano resonance originates from the destructive interference between a broad continuum resonance and a narrow discrete resonance,[1] and therefore possesses a sharp asymmetric resonant response in the spectrum.[2,3] Due to its high quality ($Q$) factor and strong dispersion in the process of light-matter interaction, Fano resonance has many potential applications such as terahertz sensing,[4] group delay,[5] and logic processing[6]. In the terahertz regime, coupled resonant structures on the platform of metamaterials are mostly utilized to generate Fano resonances. In recent years, remarkable progress has been made to increase the $Q$ factor of the Fano resonance, which requires to minimize the losses of the structures.[2] To date, structures, such as split rings, have been optimized to improve the $Q$ factor of the Fano resonance reaches up to 365 experimentally.[7,8,9] On the other hand, a tunable Fano resonance is more desirable for practical applications and has attracted many interests.[3,6,10] Different approaches combining devices with tunable structures have been reported to accomplish the tunability, including the tunability on the resonant intensity[6,11] and the central frequency[3,10]. Although great progress has been made in recent years, Fano resonances with improved $Q$ factors and flexible tunability are still desperately desired to fulfill the requirements of both experimental studies and real-world applications.

Terahertz whispering-gallery-mode resonators (WGMRs), one of the resonant structures which possess intriguing properties such as high $Q$ factors, have attracted intense interest in the terahertz regime recently. Devices with high loaded $Q$ factors up to 15000[12-15] and tunability based on Joule heating methods[16-17] are experimentally demonstrated. Moreover, Fano behavior in a single resonator with a Q factor of 1600[18] and the splitting effects in coupled quartz disks[19] are experimentally observed, without considering the tunability of these devices. These improvements on the resonant structures with high performance would offer novel solutions for the realization of tunable terahertz Fano resonance with improved *Q* factors.

In this work, we propose and demonstrate a tunable terahertz Fano resonance with a high *Q* factor of 2095 at 0.439 THz in a coupled-WGMR system. With the help of the coupled-mode theory, the scheme about demonstration and tuning of the Fano resonance is considered at first. Then a relatively low Q quartz ring and a high Q silicon ring are utilized to generate Fano resonance in the experiment. By exploiting an electrical thermo-optic tuning method, the resonant frequency of the Fano resonance can be manipulated precisely. Moreover, the resonance intensities of Fano resonances can also be engineered by adjusting the coupling strength between the two resonators. The high *Q* factor, together with the tunability, has shown great promise in designing novel terahertz devices such as terahertz sensors.

To figure out how to generate and tune the Fano resonance in the coupled-WGMR system shown in Fig. 1(a), which consists of a waveguide and two WGMRs, the coupled-mode theory is utilized to model the system. The relationships between the complex intracavity field amplitudes *A1* and *A2* can be expressed as[20,21]

$$\left(j\delta_1 + \frac{\gamma_1}{2}\right) \cdot A_1 + j\mu_1 S_i + j\mu_2 A_2 = 0 \tag{1}$$

$$\left(j\delta_2 + \frac{\gamma_2}{2}\right) \cdot A_2 + j\mu_2 A_1 = 0 \tag{2}$$

$$S_t = S_i \cdot \left(1 - \frac{\mu_1^2 \cdot \left(j\delta_2 + \frac{\gamma_2}{2}\right)}{\left(j\delta_1 + \frac{\gamma_1}{2}\right)\left(j\delta_2 + \frac{\gamma_2}{2}\right) + \mu_2^2}\right) \tag{3}$$

Where $\gamma_i$ (i=1,2) denotes the total loss of WGMR$_i$ (i=1,2). $\gamma_1=\gamma_{1i}+\gamma_c$, where $\gamma_{1i}$ refers to the intrinsic loss of WGMR1, $\gamma_c$ is the coupling loss between WGMR1 and the waveguide, and $\gamma_2=\gamma_{2i}$, where $\gamma_{2i}$ refers to the intrinsic loss of WGMR2. $\mu_i$ (i=1,2) refers to the coupling strength, modeled as $\mu_1 = \sqrt{\gamma_c}$.[18,19] $\delta_i$ (i=1,2)=$\omega$-$\omega_i$ (i=1,2) denote the detuning between the angular frequency $\omega$ of the probe terahertz waves and the resonant angular frequencies $\omega_i$. $\Delta=\omega_1-\omega_2$ denotes the angular frequency detuning between the two WGMRs. From the relations above, the complex transmitted field amplitude can be expressed as $S_t = S_i - j\sqrt{\gamma_c}A_1$, as shown as equation (3), which can be utilized to extract the intensity and phase spectra of the system.

In the assistance of equation (3), calculations are made to analyze the generation and tuning of the Fano resonance, as shown in Fig. 1(b). First of all, a relatively low $Q$ WGMR1 at zero detuning is utilized to provide a broad continuum resonance without the existence of WGMR2, and the resonance is marked as R1 in line 1. When the high $Q$ WMGR2 is coupled with WGMR1, an asymmetric Fano behavior can be observed at the position of R2, shown as line 2. Compared with line 2, only the resonant frequency of the WMGR2 is tuned and the spectrum is illustrated as line 3. In this case, the resonant frequency of the Fano resonance is also tuned. In the end, the coupling strength between two WGMRs in line 4 is decreased compared with line 3, and the resonance intensity also decreases. In summary, the Fano resonance can be realized by coupling a relatively low $Q$ resonator with a relatively high $Q$ one. The tunability of the Fano resonance can be achieved through manipulating the resonant frequency of the high Q WGMR2 and the coupling strength between two WGMRs. Moreover, to evaluate the Fano resonance, the figure of merit ($FoM=Q\times\Delta I$) parameter is considered in our experiment.[22]

In the experiment, a quartz ring with relatively low $Q$ and a silicon ring with relatively high $Q$ are utilized to be WGMR1 and WGMR2, respectively. we note that the silicon ring is made from high resistivity float zone silicon to reduce the absorption loss introduced by the material. Since the material loss of quartz is around ten times larger than that of the high resistivity float zone silicon,[12,14] it is not difficult to obtain two resonators with different Q factors. The quartz ring has an inner diameter of 4.0 mm, an outer diameter of 6.0 mm, and a thickness of 1 mm. While the silicon ring is designed to have a thickness of 0.5 mm, a 3.0 mm inner diameter and an 8.0 mm outer diameter. These parameters are designed to decrease the radiation loss of the resonator to ensure high $Q$ performances. The photograph of two WGMRs is shown in the inset of Fig. 1(a). To tune the resonant frequency of the Fano resonance, an electrical thermo-optic tuning method is applied to actively control the resonant frequency of WGMR2. A piece of resistance wire with a diameter of 0.1mm, which is made from copper, is used to form almost a circle and attached to the inside of the resonator. When the voltage is applied to the resistance wire, the temperature, the refractive index and the resonant frequency of the silicon ring would be tuned. The applied voltage is precisely controlled by a voltage source with high precision. Moreover, a silica fiber with a diameter of 0.2 mm is utilized to excite WGMs in the resonators. The silica fiber and two WGMRs are positioned by translation stages with the precision of 5 μm to adjust the coupling strength between the two WGMRs, and therefore the resonant intensity of the Fano resonance can be engineered. The coupled-WGMR system is characterized by a terahertz vector network analyzer from 0.325 to 0.5 THz and the detailed description of the setup is presented in our previous work.[13]

To actively control the Fano resonance, the electrical thermo-optic tuning process on the high Q WGMR2 is discussed at first. Here, we analyze the frequency detuning and the temperature of the ring with the change of applied voltages. Theoretically, a thermal simulation based on the finite element method and the Poisson's equation is carried out to simulate the temperature change of the silicon ring. The silicon ring is fixed by polyfoam from the inner side, and the resistance wire is attached to the top of the ring. The thermal conductivities of silicon, copper, and polyfoam are 130, 400, and 0.05 W/m·K, respectively, and the heat convection coefficient of air is set as 5 W/m$^2$·K. We assume that the electrical energy is completely converted into thermal energy in the simulation. Moreover, the frequency drift of the resonant frequency is proportional to the temperature change of the silicon ring, which can be modeled as[16,17,23]:

$$\Delta f = -\Delta T \cdot \frac{f_0}{n_c} \cdot \left( \frac{\Delta n}{\Delta T} \right) \tag{4}$$

where $\Delta n/\Delta T$ refers to the thermo-optic coefficient of the silicon ($1.37\times10^{-4}$K$^{-1}$)[16], $f_0$ is the resonant frequency of the ring, and $n_c$ is the group refractive index. Therefore, the temperature of the ring and the frequency detuning can be calculated by each other. According to the simulated results, the frequency detuning and the temperature of the ring show a linear relation with the square of the applied voltage, shown in Fig. 2(a), calculated slopes are -0.28 GHz/V$^2$ and 16.1 K/V$^2$, respectively. These two parameters refer to the efficiency of the frequency tuning and heating process. And the maximum frequency detuning and the highest temperature are -2.55 GHz and 437 K. We note that the voltage considered here is the voltage applied to the whole resistance wire, in the experiment, only 1/3 of the wire is attached to the resonator. Measured results obey an approximately linear relationship, we note that we measure the frequency drifts and calculate the temperature change with the help of equation (4). The slopes are -0.22 GHz/V$^2$ and 12.7 K/V$^2$, and the maximum values are -1.89 GHz and 400 K. In the experiment, the heat in the resistance wire would not conduct to the silicon ring in 100%, which causes the difference between measured and simulated results. Meanwhile, the temperature distribution of the silicon ring resonator is illustrated in Fig. 2(b). Considering the approximately linear relationship between the frequency detuning and the square of applied voltage, the frequency detuning can be continuously and precisely manipulated with the high-precision voltage source.

With the help of the electrical thermo-optic tuning method, tunable Fano resonance and the corresponding phase profiles are observed experimentally and shown in Fig. 3. The measured results are fitted by the aforementioned analytic model for extracting the parameters related to the state of the system. Obtained from the experimental and simulated results, Fano resonance occurs at approximately 0.439 THz. The typical bandwidth of the low $Q$ WGMR1 is $\gamma_{1i}/2\pi$=780 MHz, corresponding to an intrinsic $Q$ factor of 578. While the typical bandwidth of the high $Q$ WGMR2 is $\gamma_{2i}/2\pi$=210 MHz, corresponding to an intrinsic $Q$ factor of 2095, which is 3.6 times higher than that of the low $Q$ resonator. Meanwhile, the highest *FoM* in the tuning process is around 92 and can be observed when the voltage is 1.9V. When the voltage is lower than 1.9 V (Fig. 3 (a1) ~ (b1)), the resonant frequency of WGMR2 is larger than that of WGMR1, namely $\Delta/2\pi$<0. An asymmetric Fano behavior can be found at the higher frequency (namely right side) of Lorentzian shape resulting from the coupling between WGMR1 and WGMR2. The phase profiles corresponding to (a1) ~ (b1) are illustrated in (a2) ~ (b2). The slowly changing phase profiles below $\pi$ indicate an under-coupling state of the WGMR1, and the obvious phase change over a small frequency range is caused by WGMR2. When the voltage is 1.9V, the detuning is approximately zero ($\Delta/2\pi$≈0), a symmetric transparent window appears at the center of the Lorentz line, which is a special case of Fano resonance and often referred as electromagnetically induced transparency (EIT) behavior,[24] as illustrated in Fig. 3 (c1). The corresponding phase profile of Fig. 3 (c2), shows an abrupt change of phase shift near the transparent window in the phase change introduced by WGMR1. While the voltage is larger than 1.9 V, the resonant frequency of mode in WGMR2 is smaller than that of WGMR1 ($\Delta/2\pi$>0), Fano resonance occurs on the falling side of the low $Q$ resonance, as illustrated in Fig. 3 (d1) ~ (e1). The corresponding phase profiles are shown as (d2) ~ (e2).

Besides, the engineering of the resonance intensities is also observed in the experiment. We gradually increase the gap (corresponding to a decrease of $\mu_2/2\pi$) with the help of the transition stage and observe the variation of the spectra in both EIT and Fano occasions. For the EIT behavior (Fig. 4(a) ~ (d), U=1.9V), the spectrum of EIT behavior in Fig. 6(a) becomes the Lorentzian line of Fig. 4(d). Resonance intensities of the transparent window decrease gradually and are ~2, ~1, ~0.6, and 0 dB, respectively. Phase profiles are shown in Fig. 4 (a1) ~ (d1) and the phase shift caused by WGMR2 decreases with the increasing gap. The tuning of the Fano occasion is also observed (Fig. 4(e) ~ (h), U=1.9V), and the resonance intensities are 0.8, 0.4, 0, and 0 dB, respectively. Phase profiles of these Fano shapes are shown as Fig. 4(e1) ~ (h1). Moreover, calculated *FoM*s are listed in the figure caption of Fig. 4, where the decrease of FoMs is mainly caused by the reduction of resonance intensity during the tuning progress.

We compare the measured parameters of Fano resonance in our work, including *Q* factor, *FoM* parameter, tunability of resonant frequency, and tunability of resonance intensity, with those of reported works, as summarized in Table 1. The scheme of this work not only shows distinct improvement of Fano resonance Q factors but also introduces flexible tunability.

In summary, we propose and demonstrate a tunable Fano resonance with a high *Q* (2095) in coupled terahertz WGMRs. By introducing coupling between a relatively low *Q* (578) quartz ring resonator and a tunable high Q (2095) silicon ring resonator, Fano resonances are experimentally demonstrated at 0.439 THz. The resonant frequency of the Fano behavior can be actively controlled by an electrical thermal-optic tuning method, meanwhile, the resonance intensity can be engineered by adjusting the coupling strength between the two WGMRs. We believe our work will not only be an influential source for inspiring more novel theoretical and experimental concepts, but also further initiate the widespread use of terahertz WGMR-based devices, systems, and facilitate applications that have been proposed, but not yet implemented.

This research is supported by National Natural Science Foundation of China (NSFC) (61735006, 61905007) and Laboratory Research Fund from Wuhan National Laboratory for Optoelectronics (Grant No. 2018WNLOKF001). The authors thank the 41st research institute of China Electronics Technology Group Corporation for providing the test equipment.

**FIGURE LEGENDS**

**Fig. 1.** (a) Schematic of the proposed coupled-WGMR system, which consists of a waveguide and two coupled WGMRs. The inset illustrates the photograph of two WGMRs utilized in the experiment. (b) Calculated transmission spectra of the coupled-WGMR system in different situations according to equation (3). Parameters ($\Delta/2\pi$, $\gamma_{1i}/2\pi$, $\gamma_{2i}/2\pi$, $\gamma_c/2\pi$, $\mu_2/2\pi$) utilized in the figure are: line 1: 300 MHz, 820 MHz, 240 MHz, 470 MHz, 0 MHz; line 2: 300 MHz, 820 MHz, 240 MHz, 470 MHz, 150 MHz; line 3: 150 MHz, 820 MHz, 240 MHz, 470 MHz, 150 MHz; line 4: 150 MHz, 820 MHz, 240 MHz, 470 MHz, 150 MHz.

**Fig. 2.** (a) Relations between Frequency detuning (blue dots), temperature (red dots) of the silicon ring and the square of the applied voltage. Solid dots come from the simulated results while hollow dots are the processing of experimental results. (b) The temperature distribution of the ring structure under the applied voltage of 3V.

**Fig. 3.** (a1) ~ (e1) Transmission profiles of measured Fano resonances with the change of applied voltages on the resistance wire. The insets are zoomed parts that illustrate details of Fano resonances. (a2) ~ (e2) Observed transmission phase spectra which corresponds to (a1) ~ (e1). The insets are details of phase change introduced by Fano resonances. Blue lines refer to measured results and red lines indicate simulated lines. The frequency detuning of the two WGMRs ($\Delta/2\pi$) in (a1) ~ (e1) are -315, -105, 15, 135 and 273 MHz, respectively. $\Delta I$: Resonance intensity. *FoM*: Figure of merit parameter (*FoM*=$Q\times\Delta I$). Parameters used in the analytic model ($\Delta/2\pi$, $\gamma_{1i}/2\pi$, $\gamma_{2i}/2\pi$, $\gamma_c/2\pi$, $\mu_2/2\pi$) of (c1) are: 15 MHz, 780 MHz, 210 MHz, 350 MHz, 130 MHz.

**Fig. 4.** Measured transmission spectra in the process of tuning the resonance intensity when the applied voltage is 1.9V ((a) ~ (d)) and 1.65V ((e) ~ (h)). Corresponding transmission phase profiles are illustrated as (a1) ~ (h1). Blue lines refer to measured results and red lines indicate simulated lines. The coupling strengths ($\mu_2/2\pi$) corresponding to (a) ~ (h) are 130, 90, 75, 0, 135, 120, 95, 0 MHz, respectively. Calculated *FoM*s from (a) to (h) are 86, 36, 24, 0, 25, 15, 0, 0. The fitted parameters ($\Delta/2\pi, \gamma_{1i}/2\pi, \gamma_{2i}/2\pi, \gamma_c/2\pi, \mu_2/2\pi$) *of (a), (e) are: (a) 15 MHz, 860MHz, 200 MHz, 390 MHz, 130 MHz; (e) 200 MHz, 820 MHz, 240 MHz, 430 MHz, 135 MHz.*

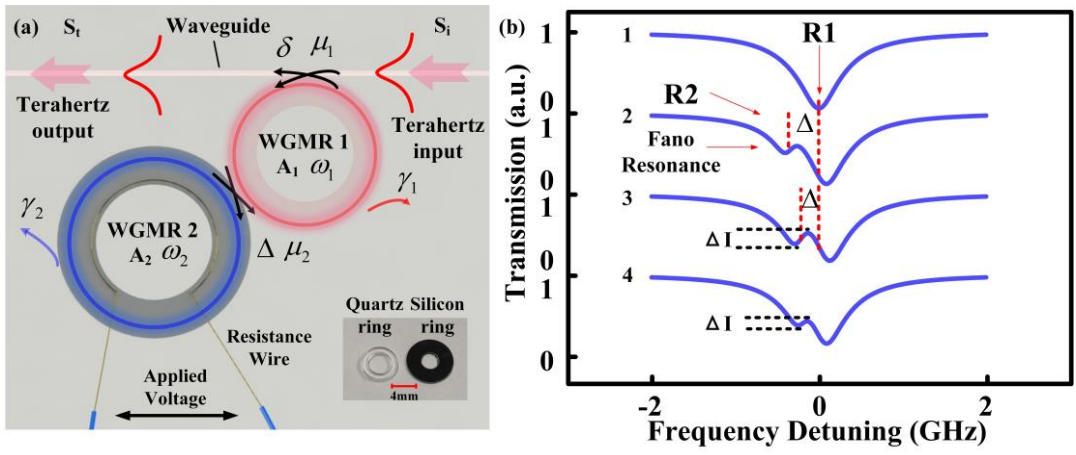

FIG. 1.

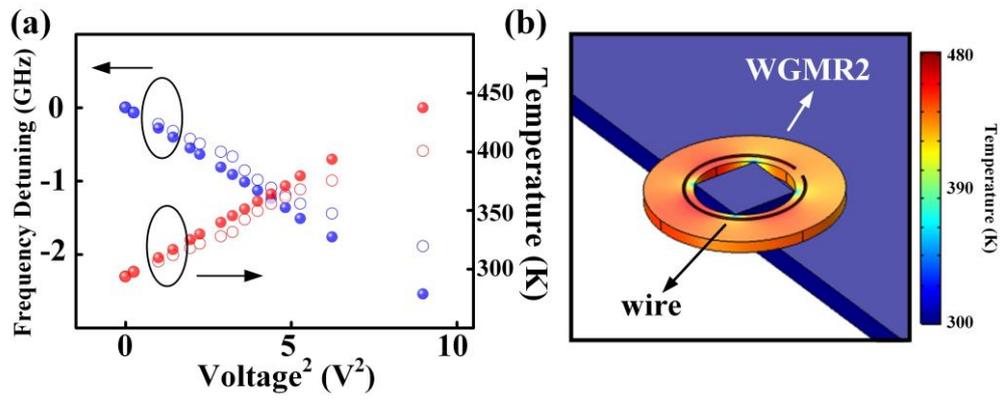

FIG. 2.

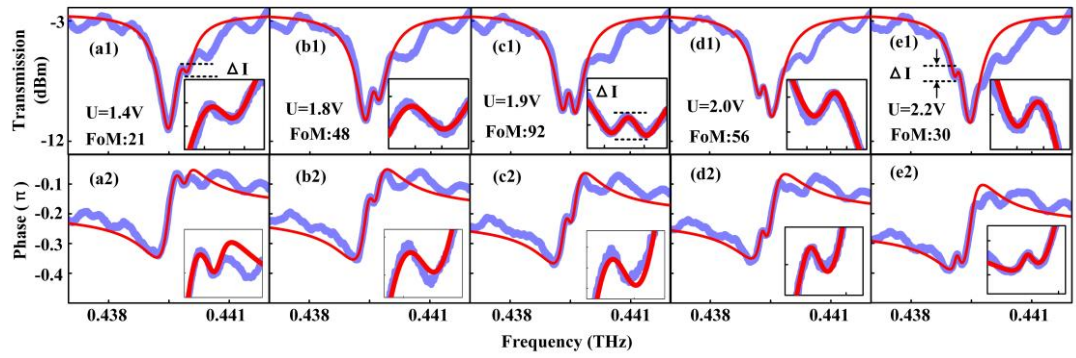

FIG. 3.

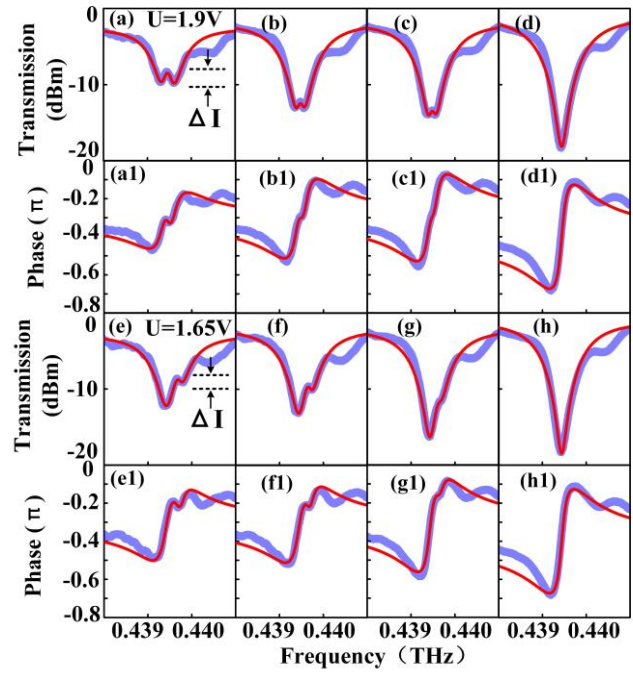

FIG. 4.

**Table 1.   Comparison of reported experimental Fano resonance Q factors and tunability.**

| Schemes | Q | FoM | Switch of line shape | Tuning of $\Delta I$ |
|---|---|---|---|---|
| 2016 (ref. 7) | 220 | 135 | N[a] | N |
| 2016 (ref. 5) | ~20 | <20 | N | Y[b] |
| 2018 (ref. 6) | 20 | <20 | N | Y |
| 2016 (ref. 8) | 167 | 7 | N | Y |
| 2012 (ref. 9) | 227 | 46 | N | N |
| 2017 (ref. 16) | 1600 | - | N | N |
| This work | 2095 | 92 | Y | Y |

[a] N refers to no; [b] Y refers to yes.

Table. 1